\documentclass{article}
\usepackage{amssymb}

\input{tcilatex}
\begin{document}

\title{ \textbf{Procedure with Massive Neutrinos for the Standard Model
Processes with Negligible Lorentz Invariance Violation}}
\author{Josip \v{S}oln \\
Army Research Laboratory (ret.),\\
JZS Phys-Tech\\
Vienna, Virginia 22182, USA\\
soln.phystech@cox.net}
\maketitle

\textbf{.}

\bigskip

\textbf{\ Abstract}

For the electroeak interactions, the massive neutrino perturbative
kinematical procedure is developed in the massive neutrino Fock space; The
perturbation expansion parameter is the ratio of neutrino mass to its
energy. This procedure, within the Pontecorvo-Maki-Nakagawa-Sakata modified
electroweak Lagrangian, calculates the cross-sections with the new neutrino
energy projection operators in the massive neutrino Fock space, resulting in
the Standard Model mass-less flavor neutrino cross-sections, plus the
Lorentz non-invariant neutrino oscillation cross-sections which are
proportional to the squares of neutrino masses and, as such, practically
unobservable in the laboratory. This scheme reinforces the notion that the
mass-less flavor neutrino can be considered as the superposition of three
massive neutrinos.

\bigskip

\textbf{\ Introduction}

The flavor changing neutrino oscillations experiments, such as, The
Super-Kamiokande [1], SNO [2], KAMLAND [3] as well as Homestake
Collaboration [4], clearly require massive neutrinos as have been exhibited,
for example, by Bilenky, Giunti and Grimus [5] , Giunti and Laveder [6] and
Kayser [7]. In discussing the neutrino oscillations, one aassumes that the
left-handed flavor mass-less neutrino fields $\nu _{\alpha L}$, with $\alpha
=e,\mu ,\tau $, are unitary linear combinations of of the massive neutrino
fields $\nu _{iL}$and analogously for the states (see [5-10])and references
therein),

\begin{equation}
\nu _{\alpha L}=U_{\alpha i}\nu _{iL},\mid \nu _{\alpha }\rangle =U_{i\alpha
}^{\dagger }\mid \nu _{i}\rangle \ \ (i=1,2,3;\alpha =e,\mu ,\tau )\ \ \ 
\text{\ \ \ }\ \ \ \   \tag{1,2}
\end{equation}

\ \ \ \ \ \ \ \ \ \ \ \ \ \ \ \ \ \ \ \ \ \ \ \ \ \ \ \ \ \ \ \ \ \ \ \ \ \
\ \ \ \ \ \ \ \ \ \ \ \ \ \ \ \ \ \ \ \ \ \ \ \ \ \ \ \ \ \ \ \ \ \ \ \ \ \
\ \ \ \ \ \ \ \ \ \ \ \ \ \ \ \ \ \ \ \ \ \ \ \ \ \ \ \ \ \ \ \ \ \ \ \ \ \
\ \ \ \ \ \ \ \ \ \ \ \ \ \ \ \ \ \ \ \ \ \ \ \ \ \ \ \ \ \ \ \ \ \ \ \ \ \
\ \ \ \ \ \ \ \ \ \ \ \ \ \ \ \ \ \ \ \ \ \ \ \ \ \ \ \ \ \ \ \ \ \ \ \ \ \
\ \ \ \ \ \ \ \ \ \ \ \ \ \ \ \ \ \ \ \ \ \ \ \ \ \ \ \ \ \ \ \ \ \ \ \ \ \
\ \ \ \ \ \ \ \ \ \ \ \ \ \ \ \ \ \ \ \ \ \ \ \ \ \ \ \ \ \ \ \ \ \ \ \ \ \
\ \ \ \ \ \ \ \ \ \ \ \ \ \ \ \ \ \ \ \ \ \ \ \ \ \ \ \ \ \ \ \ \ \ \ \ \ \
\ \ \ \ \ \ \ \ \ \ \ \ \ \ \ \ \ \ \ \ \ \ \ \ \ \ \ \ \ \ \ \ \ \ \ \ \ \
\ \ \ \ \ \ \ \ \ \ \ \ \ \ \ \ \ \ \ \ \ \ \ \ \ \ \ \ \ \ \ \ \ \ \ \ \ \
\ \ \ \ \ \ \ \ \ \ \ \ \ \ \ \ \ \ \ \ \ \ \ \ \ \ \ \ \ \ \ \ \ \ \ \ \ \
\ \ \ \ \ \ \ \ Here $U$ is the unitary Pontecorvo-Maki-Nakagawa-Sakat
(PMNS) matrix and $\nu _{iL}$\ \ is a left-handed neutrino field associated
with mass $m_{i}$ (see, for example [5-11]). In the study of flavor neutrino
oscillations, the flavor state from (2) is used, via the Schr\"{o}dinger
equation, to calculate the oscillation probability $P_{\alpha \beta }(L)$ \
for the flavor neutrino oscillation transition $\nu _{\alpha
}\longrightarrow $\ $\nu _{\beta }$\ at a very large distance $L(\backsim
thousands$ $of$ $km)$\ (see, for example [5-11] ).

However, it is well known that the standard model (SM) with mass-less flavor
neutrinos has been remarkably successful in describing the laboratory
experimental data, such as the neutrino scattering, at low and medium
energies ( see, for instance, M. Fukugita and T. Yanagida [8]; and C. Giunti
and C. W. Kim [9]). So one can then ask whether the SM cross-sections can be
derived when starting, instead with the mass-less flavor neutrino fields \ \ 
$\nu _{\alpha L}$, with the massive neutrino fields \ $\nu _{iL}$. In this
article the answer to this question is in affirmative.

To proceed in this direction, as suggested by (1), the application of the
PMNS substitution rule (3) transforms the SM Lagrangian density with the
mass-less neutrino fields into the one with the massive neutrino fields
(4):\ \ $\ $

$\ \ \ \ \ \ \ \ \ \ \ \ \ \ \ \ \ \ \ \ \ \ \ \ \ \ \ \ \ \ \ \ \ \ \ \ \ \
\ \ \ \ \ \ \ \ \ \ \ \ \ \ \ \ \ \ \ \ \ \ \ \ \ \ \ \ \ \ \ \ \ \ \ \ \ \
\ \ \ \ \ \ \ \ \ \ \ \ \ \ \ \ \ \ \ \ \ \ \ \ \ \ \ \ \ \ $%
\begin{equation}
The\text{ }PMNS\text{ }substitution:\nu _{\alpha L}\rightarrow U_{\alpha
i}\nu _{iL}  \tag{3}
\end{equation}%
\begin{eqnarray}
\alpha ,\beta ,...,\epsilon &=&e,\mu ,\tau ;i,j,a,...,b=1,2,3:  \nonumber \\
l_{\alpha L} &=&\left( 
\begin{array}{c}
U_{\alpha i}\nu _{iL} \\ 
\alpha _{L}%
\end{array}%
\right) ,\epsilon _{L,R}=P_{L,R}\epsilon ,P_{L,R}=\frac{1}{2}\left( 1\mp
\gamma ^{5}\right)  \nonumber \\
L_{W,int}^{Lepton} &=&\frac{g}{2\sqrt{2}}\dsum\limits_{\epsilon =e,\mu ,\tau
;i=1,2,3}[\overline{\nu }_{iL}(x)U_{i\epsilon }^{\dagger }\gamma ^{\mu
}\epsilon _{L}(x)W_{\mu }(x,+)  \nonumber \\
&&+\overline{\epsilon }_{L}(x)\gamma ^{\mu }U_{\epsilon j}\nu _{jL\left(
x\right) }W_{\mu }^{\dagger }(x,+)],  \TCItag{4} \\
W^{\mu }(x,\pm ) &=&\frac{1}{\sqrt{2}}\left[ W^{\mu }(x,1)\pm iW^{\mu }(x,2)%
\right]  \nonumber \\
L_{Z,int}^{Lepton} &=&\frac{g}{c_{W}}Z_{\mu }(x)\dsum\limits_{\epsilon
=e,\mu ,\tau }\left[ \overline{l}_{\epsilon L}(x)\frac{\tau _{3}}{2}\gamma
^{\mu }l_{\epsilon L}(x)-s_{W}^{2}(-)\overline{\epsilon }(x)\gamma ^{\mu
}\epsilon (x)\right]  \nonumber \\
&=&\frac{g}{4c_{W}}Z_{\mu }(x)\dsum\limits_{\epsilon =e,\mu ,\tau
,a,b=1,2,3}[\overline{\nu _{a}}(x)U_{a\epsilon }^{\dagger }\gamma ^{\mu
}\left( 1-\gamma ^{5}\right) U_{\epsilon b}\nu _{b}(x)  \nonumber \\
&&+\overline{\epsilon }(x)\gamma ^{\mu }\left[ \left( 4s_{W}^{2}-1\right)
+\gamma ^{5}\right] \epsilon (x)],  \nonumber \\
s_{W} &=&\sin \theta _{W},c_{W}=\cos \theta _{W\text{ \ \ \ \ \ \ \ \ \ \ \
\ \ \ \ \ \ \ \ \ \ \ \ \ \ \ \ \ \ \ \ \ \ \ \ \ \ \ \ \ \ \ \ \ \ \ \ \ \
\ \ \ \ \ \ \ \ \ \ \ \ \ \ \ \ \ \ }}  \nonumber
\end{eqnarray}

\ \ \ 

Since the Lagrangian densities(4) contains the massive neutrino fields, all
the calculations are now done formally in the massive neutrino Fock space.
The mass-less neutrinos will be the mass state neutrinos in the limit of
negligible masses as a result of the perturbative neutrino kinematical
procedure.

\bigskip

\textbf{Perturbative kinematical procedure for calculating the neutrino
differential cross-sections}

\bigskip

A free neutrino spinor field with the mass $m_{i}$,$i=1,2,3$ , is written
generally with the creation and annihilation operators as

\begin{eqnarray}
\nu _{i}(x) &=&\frac{1}{\left( 2\pi \right) ^{\frac{3}{2}}}\int \frac{d^{3}q%
}{q^{0}}\sum_{s}e^{iqx}u(q,s)a(q,s)+e^{-iqx}v(q,s)b^{\dagger }(q,s) 
\nonumber \\
q^{0} &=&\left( \overrightarrow{q}^{2}+m_{i}^{2}\right) ^{\frac{1}{2}}\text{
\ \ \ \ \ \ \ \ \ \ \ \ \ \ \ \ \ \ \ \ \ \ \ \ \ \ \ \ \ \ \ \ \ \ \ \ \ \
\ \ \ \ \ \ \ \ \ \ \ \ \ \ \ \ \ }  \TCItag{5}
\end{eqnarray}

The pertutbative kinematics is based on the fact that the neutrino mass \ $%
m_{i}(m_{i}\langle 1eV)$is generally much smaller than its absolute momentum
value$\left\vert \overrightarrow{q}\right\vert $. Therefore it is convenient
to start with the \textquotedblleft mass-less\textquotedblright\
four-component neutrino momentum \ $q_{(\gamma )}^{\mu }$ with fixed flavor
parameter $\gamma $

\begin{equation}
q_{\left( \gamma \right) }^{\mu }=\left( \overrightarrow{q}_{\left( \gamma
\right) },q_{\left( \gamma \right) }^{0}\right) ,q_{\left( \gamma \right)
}^{2}=0,\gamma =e,\mu ,\tau \text{ \ \ \ \ \ }  \tag{6}
\end{equation}%
Next, one assumes that under this flavor parameter \ $\gamma $ are grouped
together three massive neutrinos, say, $\nu _{i}$ with masses $m_{i}$ $%
;i=1,2,3$ then the difference among their energies $\left\vert \Delta
q_{(i_{1},i_{2})}^{0}\right\vert =\left\vert
(q_{(i_{2})}^{0}-q_{(i_{1})}^{0})\right\vert \cong \left\vert \Delta
m_{i_{2,}i_{1}}^{2}\diagup q_{(\gamma )}^{0}\right\vert =\left\vert
(m_{i_{2}}^{2}-m_{i_{1}}^{2})\diagup q_{(\gamma )}^{0}\right\vert $ is much
smaller than the quantum-mechanical uncertainty of the energy [12]. As a
consequence, in this case with fixed $\gamma $ it is impossible to
distinguish the emission of neutrinos with different masses in the neutrino
processes [12]. Hence, the three massive neutrinos, satisfying these quantum
mechanical conditions, can be viewed as superposing themselves to form the
flavor neutrino $\nu _{\gamma }$ [11,12] as depicted by relations (1) and
(2). With this in mind, with $q_{(i,\gamma )}^{\mu }$ as the four-momentum
of the massive neutrino with mass $m_{i}$ the perturbative kinematics can be
presented as

\begin{eqnarray}
q_{(i,\gamma )}^{\mu } &\simeq &q_{(\gamma )}^{\mu }+g_{0}^{\mu }\frac{%
m_{i}^{2}}{2q_{(\gamma )}^{0}},i=1,2,3;\gamma (fixed)=(e,\mu ,\tau ); 
\nonumber \\
\overrightarrow{q}_{\left( i,\gamma \right) } &=&\overrightarrow{q}_{\left(
\gamma \right) },q_{(i,\gamma )}^{0}\simeq q_{(\gamma )}^{0}+\frac{m_{i}^{2}%
}{2q_{(\gamma )}^{0}},q_{(i,\gamma )}^{2}\simeq -m_{i}^{2}\text{ \ \ \ \ \ }
\TCItag{7}
\end{eqnarray}%
In (7) the terms with $O(m_{i}^{4})$ have been neglected and the fixed
parameter $\gamma $, as already established is the neutrino flavor. Taking
these relations into account, within the massive neutrino Fock space the
differential cross-sections with flavor neutrinos are calculated. The
question, of course is: is the result consistent with the SM?

To continue, in analogy to $q_{(i,\gamma )}^{\mu }$, one now introduces $%
\widehat{s}_{(i,\gamma )}$ and $s_{(i,\gamma )}$ to denote respectively, the
helicity operators and eigenvalues for \ $i=1,2,3$ massive neutrinos
comprising the mass-less flavor neutrino $\nu _{\gamma }$; The helicity
operator and eigenvalue of the mass-less flavor neutrino $\ \nu _{\gamma }$
are denoted, respectively, as $\widehat{s}_{(\gamma )}$ and $s_{(\gamma )%
\text{ }}$. And, the effects of the massive to mass-less-neutrino
kinematical relation (7) on these helicity eigenvalues are simply, what one
can call, the ordinary massive to mass-less neutrino helicity relation.

\begin{eqnarray}
\widehat{s}_{(i,\gamma )} &=&\overrightarrow{q}_{\left( i,\gamma \right)
}\cdot \overrightarrow{\sigma }\diagup \overrightarrow{\downharpoonright q}%
_{\left( i,\gamma \right) }\downharpoonright =\overrightarrow{q}_{\left(
\gamma \right) }\cdot \overrightarrow{\sigma }\diagup \overrightarrow{%
\downharpoonright q}_{\left( \gamma \right) }\downharpoonright =\widehat{s}%
_{(\gamma )},  \nonumber \\
\widehat{s}_{(i,\gamma )} &=&\widehat{s}_{(\gamma )}=\widehat{s}_{(k,\gamma
)}\Longrightarrow s_{(i,\gamma )}=s_{(\gamma )}=s_{(k,\gamma )},etc; 
\TCItag{8} \\
i\text{ }or\text{ }k,... &=&1,2,3;\gamma (fixed)=(e\text{ }or\text{ }\mu 
\text{ }or\text{ }\tau )  \nonumber
\end{eqnarray}

As a consequence of (7) and (8), with spinor indices suppressed, the
contractions of massive neutrino free-field operators with the massive
neutrino states are

\begin{eqnarray}
\left\langle 0\right\vert \text{\ }\nu \left( x,l\right) \left\vert
q_{\left( i,\gamma \right) },s_{\left( i,\gamma \right) }\right\rangle \text{%
\ } &=&\frac{1}{\left( 2\pi \right) ^{\frac{3}{2}}}\text{\ }e^{i\left(
q_{\left( i,\gamma \right) }\cdot x\right) }\delta _{li}u\left( q_{\left(
i,\gamma \right) },s_{_{\left( i,\gamma \right) }}\right) ,\text{\ \ \ \ \ }
\nonumber \\
\left\langle q_{\left( j,\delta \right) },s_{\left( j.\delta \right)
}\right\vert \text{\ }\overline{\nu }\left( x,k\right) \text{\ }\left\vert
0\right\rangle  &=&\frac{1}{\left( 2\pi \right) ^{\frac{3}{2}}}\text{\ }%
e^{i\left( q_{\left( k,\delta \right) }\cdot x\right) }\delta _{kj}\overline{%
v}\left( q_{\left( j,\delta \right) },s_{_{\left( j,\delta \right) }}\right) 
\text{\ \ \ \ \ \ \ \ \ \ \ \ \ \ \ \ \ \ \ \ \ \ \ \ }  \TCItag{9}
\end{eqnarray}%
where $s_{(j,\delta )}$ and $\delta $ have the same kind interrelationship \
as $s_{(i,\gamma )}$ and $\gamma $ in (8), etc.

Since , as shown in (7) to (9), the superposed three massive neutrinos
contain the single flavor designation, either in the initial or final state,
say, $\gamma $ and $\delta $, the process can be denoted as $\ \nu (\gamma
)+\alpha (P_{1})\rightarrow \nu (\delta )+\beta (P_{2})$ .From the
Lagrangian densities (4) the amplitude and its Hermitian conjugate for the
process containing massive neutrinos, are build around these respective
flavor designations, $\gamma $ and $\delta $. so that the generic amplitudes
are given, respectively, as

\begin{eqnarray}
S_{amp} &\backsim &\sum_{i,j,...}\delta _{4}(q_{(i,\gamma
)}+P_{(1)}-q_{(j,\delta )}-P_{(2)})iM_{i,j,...},  \nonumber \\
S_{amp}^{\dagger } &\backsim &\sum_{k,l,...}\delta _{4}(q_{(k,\gamma
)}+P_{(1)}-q_{(l,\delta )}-P_{(2)})(-i)M_{k,l,...}^{\ast }  \TCItag{10}
\end{eqnarray}%
Here, the momenta indicate the actual massive neutrino-lepton scattering and
different Latin indices indicate possibilities of summation with the $U$
matrices which, however, here is not necessary to be explicit. To derive the
cross-section, with the help of (7), (8) and (9), one needs

\begin{eqnarray}
S_{amp}^{\ast }S_{amp} &=&\sum_{i.j,...;k,l,...}\left\{ \delta
_{4}^{2}(q_{(\gamma )}+P_{(1)}-q_{(\delta )}-P_{(2)})\right.  \nonumber \\
&&+\delta _{3}^{2}(\overrightarrow{q}_{(\gamma )}+\overrightarrow{P}_{(1)}-%
\overrightarrow{q}_{(\delta )}-\overrightarrow{P}_{(2)})\delta (q_{(\gamma
)}^{0}+P_{(1)}^{0}-q_{(\delta )}^{0}-P_{(2)}^{0})  \nonumber \\
&&\times \delta \prime (q_{(\gamma )}^{0}+P_{(1)}^{0}-q_{(\delta
)}^{0}-P_{(2)}^{0})\frac{1}{2}\left[ \frac{m_{i}^{2}+m_{k}^{2}}{q_{(\gamma
)}^{0}}-\frac{m_{j}^{2}+m_{l}^{2}}{q_{(\delta )}^{0}}\right]  \TCItag{11} \\
&&\left. +O(m^{4})\right\} (M_{i,j,...})(M_{k,l,...}^{\ast })  \nonumber \\
&=&\delta _{4}^{2}(q_{(\gamma )}+P_{(1)}-q_{(\delta
)}-P_{(2)})\sum_{i,j,...;k,l,...}(M_{i,j,...})(M_{k,l,...}^{\ast })+O\left(
m^{4}\right) \text{\ }  \nonumber
\end{eqnarray}%
The final result in (11) is the consequence of general delta function
property $\delta (x)\delta ^{\prime }(x)=0$ \ .The terms with $O\left(
m^{4}\right) $ , denoting the fourth power of products of variety of $%
m_{i},m_{k},etc.$, are neglected. It follows that while the Fock space
contains the massive neutrino states, the cross-section will utilize the
kinematics of massless flavor neutrinos.

Next, one needs the spinor expressions, appearing in (9), to reflect
respectively, the kinematical and helicity relations in order to facilitate
the cross-section calculations.%
\begin{eqnarray}
u(q_{(i,\alpha )},s_{(i,\alpha )}) &=&\frac{m_{i}-\underline{q}_{(i,\alpha )}%
}{\sqrt{2\left( m_{i}+q_{(i,\alpha )}^{0}\right) }}u(m_{i},\overrightarrow{0}%
,s_{(i,\alpha )}),  \nonumber \\
u(m_{i},\overrightarrow{0},s_{(i,\alpha )} &=&\pm 1)=\left( 
\begin{array}{c}
1 \\ 
0 \\ 
0 \\ 
0%
\end{array}%
\right) ,\left( 
\begin{array}{c}
0 \\ 
1 \\ 
0 \\ 
0%
\end{array}%
\right) ,  \TCItag{12} \\
\underline{q}_{(i,\alpha )} &=&\gamma _{\mu }q_{(i,\alpha )};  \nonumber \\
\overline{u}(q_{(i,\alpha )},s_{(i,\alpha )}) &=&\overline{u}(m_{i},%
\overrightarrow{0},s_{(i,\alpha )})\frac{m_{i}-\underline{q}_{(i,\alpha )}}{%
\sqrt{2\left( m_{i}+q_{(i,\alpha )}^{0}\right) }}  \nonumber
\end{eqnarray}

For a process $\ $with $\gamma $ and $\delta $ flavor designations, $\ \nu
(\gamma )+\alpha (P_{1})\longrightarrow \nu (\beta )+\beta (P_{2})$, in
cross-section evaluations, one will deal with the neutrino energy projection
operator over the positive energy states. Furthermore, rather than averaging
over, one simply sums over the massive neutrino helicity degrees of freedom.
Consistent with the ordinary neutrino helicity relation (8), the sum is
carried over only the equal helicity eigenvalues:

\begin{eqnarray}
s_{(i,\alpha )} &=&s_{(k,\alpha )}=s_{(\alpha )}:\sum_{s_{(i,\alpha
)},s_{(k,\alpha )}}u(q_{(i,\alpha )},s_{(i,\alpha )})\otimes \overline{u}%
(q_{(k,\alpha )},s_{(k,\alpha )})  \nonumber \\
&=&\sum_{s_{(\alpha )}}u(q_{(i,\alpha )},s_{(\alpha )})\otimes \overline{u}%
(q_{(k,\alpha )},s_{(\alpha )})\equiv \frac{1}{2}\left[ q_{(i,\alpha
)},q_{(k,\alpha )};+,c\right] ,  \nonumber \\
\left[ q_{(i,\alpha )},q_{(k,\alpha )};+,c\right] &=&\frac{\left( m_{i}-%
\underline{q}_{(i,\alpha )}\right) \left( 1+\gamma ^{0}\right) \left( m_{k}-%
\underline{q}_{(k,\alpha )}\right) }{2\left[ \left( m_{i}+q_{(i,\alpha
)}^{0}\right) \left( m_{k}+q_{(k,\alpha )}^{0}\right) \right] ^{\frac{1}{2}}}%
,  \TCItag{13} \\
i\text{ }and\text{ }k &=&1,2,3;\text{ }\gamma =e\text{ }or\text{ }\mu \text{ 
}or\text{ }\tau  \nonumber
\end{eqnarray}%
where the $+$ sign refers to the positive energy states and $c$ refers to
the fact that the equal helicity eigenvalues in the sum yield the coherent
result. (The incoherent projection operators \ $\left[ q_{(i,\alpha
)},q_{(k,\alpha )};+,i\right] $ with unequal helicity eigenvalues \ \ $%
s_{(i,\alpha )}\neq s_{(k,\alpha )}$\ \ \ are not dealt here.) The relation
(13) defines the spinorial massive neutrino to mass-less neutrino helicity
relation and it is consistent with the ordinary helicity relation (8).
Carrying out the indicated operations in relation (13) as a power series
over the neutrino masses /energy, one obtains for the neutrino energy
projection operator over the positive energy states the following

\begin{eqnarray}
\left[ q_{(i,\alpha )},q_{(k,\alpha )};+,c\right] &=&\sum_{n=0}^{2}\left[
q_{(i,\alpha )},q_{(k,\alpha )};+,c\right] _{n},  \nonumber \\
\left[ q_{(i,\alpha )},q_{(k,\alpha )};+,c\right] _{0} &=&-\underline{q}%
_{\left( \alpha \right) },  \TCItag{14} \\
\left[ q_{(i,\alpha )},q_{(k,\alpha )};+,c\right] _{1} &=&m_{k}+\frac{\left(
m_{k}-m_{i}\right) \gamma ^{0}\underline{q}_{\left( \alpha \right) }}{%
2q_{\left( \alpha \right) }^{0}},  \nonumber \\
\left[ q_{(i,\alpha )},q_{(k,\alpha )};+,c\right] _{2} &=&-\frac{\left(
m_{k}-m_{i}\right) ^{2}\underline{q}_{\left( \alpha \right) }}{8q_{\left(
\alpha \right) }^{02}}+\frac{m_{i}m_{k}\gamma ^{0}}{2q_{\left( \alpha
\right) }^{0}}  \nonumber
\end{eqnarray}

The coherent energy operator \ $\left[ q_{(i,\alpha )},q_{(k,\alpha )};+,c%
\right] $ generates the electroweak interactions that are the same as the SM
interactions plus the LIV neutrino oscillation processes that are negligible
since their cross-sections are proportional to the squares of neutrino
masses and, as such, are essentially zero (LIV = Lorentz invariance
violating and LI = Lorentz invariant) . Relation (14) is in essence the
procedure for calculating the cross-sections for the processes requiring
only the neutrino energy projection operators over the positive energy
states.

\bigskip

\textbf{Applications to the differential cross-section calculations}

\bigskip

As established earlier and consistent with (11), the quasi-elastic
electroweak process with massive neutrinos present, to $O(m^{2})$, can can
be denoted with the kinematics that uses just the mass-less flavor neutrinos.

\begin{equation}
\nu \left( q_{\left( \gamma \right) }\right) +\alpha \left( P_{\left(
1\right) }\right) \longrightarrow \nu \left( q_{\left( \delta \right)
}\right) +\beta \left( P_{\left( 2\right) }\right) ;y=\frac{q_{\left( \gamma
\right) }^{0}-q_{\left( \delta \right) }^{0}}{q_{\left( \gamma \right) }^{0}}%
=\frac{P_{\left( 2\right) }^{0}-P_{\left( 1\right) }^{0}}{q_{\left( \gamma
\right) }^{0}}  \tag{15}
\end{equation}%
where $y$ is the momentum transfer. Now, although working in the massive
neutrino Fock space, relation (11) says that the kinematics for the
cross-sections for the quasi-elastic scattering of the massless flavor
neutrinos is determined with flavor neutrino momenta according to $\delta
_{4}\left( q_{\left( \gamma \right) }+P_{\left( 1\right) }-q_{\left( \delta
\right) }-P_{\left( 2\right) }\right) $ . Furthermore, since (see also [8])

\begin{equation}
\int dy=\frac{1}{2\pi }\int d\sigma \left( q_{\left( \delta \right) }\right)
d\sigma \left( P_{\left( 2\right) }\right) \delta _{4}\left( q_{\left(
\gamma \right) }+P_{\left( 1\right) }-q_{\left( \delta \right) }-P_{\left(
2\right) }\right) ,d\sigma \left( q\right) =\left( \frac{d^{3}q}{q^{0}}%
\right)  \tag{16}
\end{equation}%
the normalized neutrino energy transfer \ $y=\left( q_{\left( \gamma \right)
}^{0}-q_{\left( \delta \right) }^{0}\right) \diagup q_{\left( \gamma \right)
}^{0}$ \ cannot affect Lorentz invariance of any of the differential
cross-sections.

Also, in view of (11), the cross-section normalization factor is defined
with respect to the massless flavor neutrino momenta.

\[
B=\frac{1}{\left( 2\pi \right) ^{6}}\left\vert \left( P_{\left( 1\right)
}\cdot q_{\left( \gamma \right) }\right) ^{2}-P_{\left( 1\right)
}^{2}q_{\left( \gamma \right) }^{2}\right\vert ^{\frac{1}{2}}=\frac{1}{%
\left( 2\pi \right) ^{6}}\left\vert \left( P_{\left( 1\right) }\cdot
q_{\left( \gamma \right) }\right) \right\vert 
\]%
In explicit evaluations, one uses the following short-hand notations:

\begin{equation}
m_{\alpha \beta }=\sum_{i}U_{\alpha i}m_{i}U_{i\beta }^{\dagger },m_{\alpha
\beta }^{2}=\sum_{i}U_{\alpha i}m_{i}^{2}U_{i\beta }^{\dagger }\text{ \ \ \
\ \ \ \ \ \ \ \ \ \ \ \ }  \tag{17}
\end{equation}

\bigskip

\textbf{\ Deriving the differential cross-sections with new energy
projection operators for the flavor neutrino processes within the massive
neutrino Fock space----}

\bigskip

$\frac{d\sigma _{W}}{dy}-$From the Lagrangian density in (4), the free
neutrino field (5), the kinematical relation (7), the relations (8) and (9),
one derives in the usual way the \ $W-$exchange $S_{W}$ and $S_{W}^{\dagger
} $ matrix elements for the process in (15). Specifically, with the Fierz
rearrangement and repeated indices summing up, one has,

\begin{eqnarray}
S_{W} &=&\sum_{i,j}\delta _{4}\left( q_{\left( i,\gamma \right) }+P_{\left(
1\right) }-q_{\left( j,\delta \right) }-P_{\left( 2\right) }\right) \delta
_{\alpha \beta }U_{\delta j}U_{i\gamma }^{\dagger }U_{j\beta }^{\dagger
}U_{\alpha i}\frac{ig^{2}}{\left( 2\pi \right) ^{2}8M_{W}^{2}}  \nonumber \\
&&\times \overline{u}\left( q_{\left( j,\delta \right) },s_{\left( j,\delta
\right) }\right) \gamma ^{\mu }\left( 1-\gamma ^{5}\right) u\left( q_{\left(
i,\gamma \right) },s_{\left( i,\gamma \right) }\right) \overline{u}\left(
P_{\left( 2\right) },r_{2}\right) \gamma _{\mu }u\left( P_{\left( 1\right)
,}r_{1}\right) \text{\ \ \ \ \ \ \ \ \ \ \ \ \ \ \ }  \TCItag{18}
\end{eqnarray}%
and $\ S_{W}^{\dagger }$ \ \ is obtained from (18) as shown in (10). The
contribution to the process (15) due to the $W-$exchange from (18), after
taking into account (11), (17), $\sqrt{2}g^{2}=8M_{W}^{2}G$, and the fact
that \ $s_{\left( i,\gamma \right) }$, $s_{\left( j,\delta \right) }$, ...,
obey, respectively, the ordinary and spinorial helicity relation, (8) and
(13), the standard procedure gives,

\begin{eqnarray}
s_{\left( i,\gamma \right) } &=&s_{\left( g,\gamma \right) }=s_{\left(
\gamma \right) };s_{\left( h,\delta \right) }=s_{\left( j,\delta \right)
}=s_{\left( \delta \right) }:  \nonumber \\
\frac{d\sigma _{W}\left( m\right) }{dy} &=&\frac{d\sigma _{W}^{\left(
c,c\right) }\left( m\right) }{dy}=\frac{G^{2}}{4\pi \left\vert \left(
P_{\left( 1\right) }\cdot q_{\left( \gamma \right) }\right) \right\vert }%
\frac{\delta _{\alpha \beta }}{2^{5}}\sum_{i,j;g,h}\left( U_{h\delta
}^{\dagger }U_{\gamma g}U_{\beta h}U_{g\alpha }^{\dagger }\right) \left(
U_{\delta j}U_{i\gamma }^{\dagger }U_{j\beta }^{\dagger }U_{\alpha i}\right)
\nonumber \\
&&\times \left[ Tr\left( M_{1}-\underline{P}_{\left( 1\right) }\right)
\gamma _{\nu }\left( 1-\gamma ^{5}\right) \left( M_{2}-\underline{P}_{\left(
2\right) }\right) \gamma _{\mu }\left( 1-\gamma ^{5}\right) \right] 
\TCItag{19} \\
&&\times \left[ Tr\left[ q_{\left( i,\gamma \right) },q_{\left( g,\gamma
\right) };+,c\right] \gamma ^{\nu }\left( 1-\gamma ^{5}\right) \left[
q_{\left( h,\delta \right) },q_{\left( j,\delta \right) };+,c\right] \gamma
^{\mu }\left( 1-\gamma ^{5}\right) \right]  \nonumber
\end{eqnarray}%
where $m$ symbolically denotes dependence on $m_{1,2,3\text{ . }}$Next, the
coherent energy operator expansion according to (14), with gamma matrices
traces carried out, yields

\begin{eqnarray}
\frac{d\sigma _{W}\left( m\right) }{dy} &=&\frac{d\sigma _{W}\left(
SM\right) }{dy}\left[ 1+\frac{m_{\alpha \alpha }^{2}}{4}\left( \frac{1}{%
q_{\left( \gamma \right) }^{02}}+\frac{1}{q_{\left( \delta \right) }^{02}}%
\right) \right] -\frac{2G^{2}\delta _{\alpha \beta }}{\pi \left\vert \left(
P_{\left( 1\right) }\cdot q_{\left( \gamma \right) }\right) \right\vert } 
\nonumber \\
&&\times \left\{ \frac{\delta _{\alpha \gamma }}{4}m_{\alpha \delta
}m_{\delta \alpha }\left[ \frac{\left( P_{\left( 1\right) }\cdot q_{\left(
\gamma \right) }\right) \left( P_{\left( 2\right) }\cdot q_{\left( \delta
\right) }\right) }{q_{\left( \delta \right) }^{02}}+\frac{2P_{\left(
2\right) }^{0}\left( P_{\left( 1\right) }\cdot q_{\left( \gamma \right)
}\right) }{q_{\left( \delta \right) }^{0}}\right] \right.  \TCItag{20} \\
&&\left. +\frac{\delta _{\beta \delta }}{4}m_{\alpha \gamma }m_{\gamma
\alpha }\left[ \frac{\left( P_{\left( 1\right) }\cdot q_{\left( \gamma
\right) }\right) \left( P_{\left( 2\right) }\cdot q_{\left( \delta \right)
}\right) }{q_{\left( \gamma \right) }^{02}}+\frac{2P_{\left( 1\right)
}^{0}\left( P_{\left( 2\right) }\cdot q_{\left( \delta \right) }\right) }{%
q_{\left( \gamma \right) }^{0}}\right] \right\} +O(m^{4}),  \nonumber \\
\frac{d\sigma _{W}\left( SM\right) }{dy} &=&\frac{2G^{2}\delta _{\alpha
\beta }\delta _{\alpha \gamma }\delta _{\beta \delta }}{\pi \left\vert
\left( P_{\left( 1\right) }\cdot q_{\left( \gamma \right) }\right)
\right\vert }\left( P_{\left( 1\right) }\cdot q_{\left( \gamma \right)
}\right) \left( P_{\left( 2\right) }\cdot q_{\left( \delta \right) }\right) 
\nonumber
\end{eqnarray}%
One can notice that , while the negligible $LIV$ is associated with the
neutrino mass, the $LI$ Standard Model result is formally identified with
zero neutrino mass limits

\begin{equation}
\frac{d\sigma _{W}\left( m\right) }{dy}=\frac{d\sigma _{W}\left( SM\right) }{%
dy}+O\left( m^{2};LIV\right)  \tag{21}
\end{equation}%
one can summarize the neutrino flavor transitions for the $W-$exchange
neutrino processes. Flavor conserving are: $LI$ to $O(m=0)$ terms and
negligible $LIV$ to $O(m^{2})$ terms. Flavor violating are: negligible $LIV$
to $O(m^{2})$ terms.

\bigskip

$\frac{d\sigma _{Z}}{dy}-$As in the previous case, from the Lagrangian
density in (4), the free neutrino field (5), the kinematical relation (7),
the contractions (9), one derives in the usual way the $Z-exchange$ $S_{Z}$
and $S_{Z}^{\dagger }$ matrix elements for the process in (15).
Specifically, one has

\begin{eqnarray}
S_{Z} &=&\sum_{i,j}\delta _{4}\left( q_{\left( i,\gamma \right) }+P_{\left(
1\right) }-q_{\left( j,\delta \right) }-P_{\left( 2\right) }\right) \delta
_{ij}\delta _{\alpha \beta }U_{\delta j}U_{i\gamma }^{\dagger }  \nonumber \\
&&\times \frac{ig^{2}}{\left( 2\pi \right) ^{2}16c_{W}^{2}M_{Z}^{2}}\left[ 
\overline{u}\left( q_{\left( j,\delta \right) },s_{\left( j,\delta \right)
}\right) \gamma ^{\mu }\left( 1-\gamma ^{5}\right) u\left( q_{\left(
i,\gamma \right) },s_{\left( i,\gamma \right) }\right) \right.  \TCItag{22}
\\
&&\left. \times \overline{u}\left( P_{\left( 2\right) },r_{2}\right) \gamma
_{\mu }\left( 4s_{W}^{2}-1+\gamma ^{5}\right) u\left( P_{\left( 1\right)
,}r_{1}\right) \right]  \nonumber
\end{eqnarray}%
while \ $S_{Z}^{\dagger }$ is obtained from the $S_{Z}$ through the
Hermitian conjugation. \ In what follows, one will find the following
shorthand notation very useful:

\begin{eqnarray}
w_{0} &=&s_{W}^{2},w_{1}=2s_{W}^{2}-1,z_{1}=s_{W}^{2}\left(
2s_{W}^{2}-1\right) +\frac{1}{4},z_{2}=s_{W}^{2}\left( 2s_{W}^{2}-1\right) ,
\nonumber \\
z_{3} &=&s_{W}^{2}-\frac{1}{4},z_{4}=s_{W}^{2}\left( s_{W}^{2}-1\right) +%
\frac{1}{4},z_{1}+z_{3}=2s_{W}^{4}  \TCItag{23}
\end{eqnarray}

After taking into account that $c_{W}^{2}M_{Z}^{2}=M_{W}^{2}$ and the fact
that the helicities, $s_{\left( i,\gamma \right) },s_{\left( j,\delta
\right) },...$, obey both the ordinary and the spinorial helicity relations
(8) and (13), the standard procedure yields the general expressions:

\begin{eqnarray}
s_{\left( i,\gamma \right) } &=&s_{\left( k,\gamma \right) }=s_{\left(
\gamma \right) };s_{\left( l,\delta \right) }=s_{\left( j,\delta \right)
}=s_{\left( \delta \right) }:  \nonumber \\
\frac{d\sigma _{Z}\left( m\right) }{dy} &=&\frac{d\sigma _{Z}^{\left(
c,c\right) }\left( m\right) }{dy}=\frac{G^{2}}{4\pi \left\vert \left(
P_{\left( 1\right) \cdot }q_{\left( \gamma \right) }\right) \right\vert }%
\frac{\delta _{\alpha \beta }}{2^{5}}\sum_{i,j;k,l}\left( U_{\delta
j}U_{i\gamma }^{\dagger }U_{l\delta }^{\dagger }U_{\gamma k}\right) \delta
_{ij}\delta _{kl}  \TCItag{24} \\
&&\times \left[ Tr\left( M_{1}-\underline{P}_{\left( 1\right) }\right)
\gamma _{\nu }\left( 2z_{3}+\frac{1}{2}\gamma ^{5}\right) \left( M_{2}-%
\underline{P}_{\left( 2\right) }\right) \gamma _{\mu }\left( 2z_{3}+\frac{1}{%
2}\gamma ^{5}\right) \right]  \nonumber \\
&&\times \left[ Tr\left[ q_{\left( i,\gamma \right) },q_{\left( k,\gamma
\right) };+,c\right] \gamma ^{\nu }\left( 1-\gamma ^{5}\right) \left[
q_{\left( l,\delta \right) },q_{\left( j,\delta \right) };+,c\right] \gamma
^{\mu }\left( 1-\gamma ^{5}\right) \right]  \nonumber
\end{eqnarray}%
The coherent energy operator expansion according to (14), with evaluating
the traces of gamma matrices, yields

\begin{eqnarray}
\frac{d\sigma _{Z}\left( m\right) }{dy} &=&\frac{d\sigma _{Z}\left(
SM\right) }{dy}\left[ 1+\frac{m_{\gamma \gamma }^{2}}{4}\left( \frac{1}{%
q_{\left( \gamma \right) }^{02}}+\frac{1}{q_{\left( \delta \right) }^{02}}%
\right) \right] -\frac{G^{2}m_{\gamma \delta }m_{\delta \gamma }\delta
_{\alpha \beta }}{8\pi \left\vert \left( P_{\left( 1\right) \cdot }q_{\left(
\gamma \right) }\right) \right\vert }  \nonumber \\
&&\times \left\{ 2\left( \frac{1}{q_{\left( \gamma \right) }^{02}}+\frac{1}{%
q_{\left( \delta \right) }^{02}}\right) \left[ M_{1}M_{2}z_{2}\left(
q_{\left( \gamma \right) }\cdot q_{\left( \delta \right) }\right) \right.
\right.  \nonumber \\
&&+\left( P_{\left( 1\right) }\cdot q_{\left( \delta \right) }\right) \left(
P_{\left( 2\right) }\cdot q_{\left( \gamma \right) }\right) \left(
z_{1}+z_{3}\right)  \nonumber \\
&&\left. +\left( P_{\left( 1\right) }\cdot q_{\left( \gamma \right) }\right)
\left( P_{\left( 2\right) }\cdot q_{\left( \delta \right) }\right) \left(
z_{1}-z_{3}\right) \right]  \nonumber \\
&&+\frac{4}{q_{\left( \delta \right) }^{0}}\left[ M_{1}M_{2}z_{2}q_{\left(
\gamma \right) }^{0}+P_{\left( 1\right) }^{0}\left( P_{\left( 2\right)
}\cdot q_{\left( \gamma \right) }\right) \left( z_{1}+z_{3}\right) \right. 
\TCItag{25} \\
&&\left. +P_{\left( 2\right) }^{0}\left( P_{\left( 1\right) }\cdot q_{\left(
\gamma \right) }\right) \left( z_{1}-z_{3}\right) \right] +\frac{4}{%
q_{\left( \gamma \right) }^{0}}\left[ M_{1}M_{2}z_{2}q_{\left( \delta
\right) }^{0}\right.  \nonumber \\
&&\left. \left. +P_{\left( 2\right) }^{0}\left( P_{\left( 1\right) }\cdot
q_{\left( \delta \right) }\right) \left( z_{1}+z_{3}\right) +P_{\left(
1\right) }^{0}\left( P_{\left( 2\right) }\cdot q_{\left( \delta \right)
}\right) \left( z_{1}-z_{3}\right) \right] \right\}  \nonumber \\
&&+O(m^{4})\text{\ }  \nonumber \\
\frac{d\sigma _{Z}\left( SM\right) }{dy} &=&\frac{G^{2}\delta _{\alpha \beta
}\delta _{\gamma \delta }}{\pi \left\vert \left( P_{\left( 1\right) \cdot
}q_{\left( \gamma \right) }\right) \right\vert }  \nonumber \\
&&\times \left[ M_{1}M_{2}z_{2}\left( q_{\left( \gamma \right) }\cdot
q_{\left( \delta \right) }\right) +\left( P_{\left( 1\right) }\cdot
q_{\left( \delta \right) }\right) \left( P_{\left( 2\right) }\cdot q_{\left(
\gamma \right) }\right) \left( z_{1}+z_{3}\right) \right. \text{ }  \nonumber
\\
&&\left. +\left( P_{\left( 1\right) }\cdot q_{\left( \gamma \right) }\right)
\left( P_{\left( 2\right) }\cdot q_{\left( \delta \right) }\right) \left(
z_{1}-z_{3}\right) \right] \text{\ \ \ \ \ \ \ \ \ \ \ \ \ \ \ \ \ \ \ \ \ \
\ \ \ \ \ \ \ \ \ \ \ \ \ \ \ \ \ \ \ \ \ \ \ \ \ \ \ \ \ \ \ \ \ \ \ \ \ \
\ \ \ \ \ \ \ \ \ \ \ \ }  \nonumber
\end{eqnarray}%
Here also,the negligible $LIV$ is associated with the neutrino mass while
the $LI$ Standard Model result is identified with formally zero neutrino
mass limits :

\begin{equation}
\frac{d\sigma _{Z}\left( m\right) }{dy}=\frac{d\sigma _{Z}\left( SM\right) }{%
dy}+O\left( m^{2};LIV\right) ,  \tag{26}
\end{equation}%
While the terms of $O\left( m=0\right) $ are $LI$ \ and flavor conserving,
the negligible $LIV$ terms \ of $O(m^{2})$are either flavor violating or
flavor conserving.

\bigskip

$\frac{d\sigma _{\left\{ W,Z\right\} }}{dy}-$Here, the differential
cross-section for the quasi-elastic neutrino scattering (15) due to the
overlapping $S-matrix$ elements from the $W-$ and $Z-$is given as a sum of
its components after taking into account relations (11), (18), and (22).
Importantly, again taking into account the fact that helicities, $s_{\left(
i,\gamma \right) ,s_{\left( j,\delta \right) }},...,$ obey both the ordinary
and spinorial helicity relations, (8) and (13), the standard procedure
yields the general expression

\begin{eqnarray}
s_{\left( e,\gamma \right) } &=&s_{\left( g,\gamma \right) }=s_{\left(
i,\gamma \right) }=s_{\left( k,\gamma \right) }=s_{\left( \gamma \right) }; 
\nonumber \\
s_{\left( h,\delta \right) } &=&s_{\left( f,\delta \right) }=s_{\left(
l,\delta \right) }=s_{\left( j,\delta \right) }=s_{\left( \delta \right) }: 
\nonumber \\
\frac{d\sigma _{\left\{ W,Z\right\} }\left( m\right) }{dy} &=&\frac{d\sigma
_{\left\{ W,Z\right\} }^{\left( c,c\right) }\left( m\right) }{dy}=\frac{G^{2}%
}{8\pi \left\vert \left( P_{\left( 1\right) \cdot }q_{\left( \gamma \right)
}\right) \right\vert }\frac{\delta _{\alpha \beta }}{2^{5}}  \nonumber \\
&&\times \left\{ \left[ Tr\left( M_{1}-\underline{P}_{\left( 1\right)
}\right) \gamma _{\nu }\left( 4z_{3}+\gamma ^{5}\right) \left( M_{2}-%
\underline{P}_{\left( 2\right) }\right) \gamma _{\mu }\left( 1-\gamma
^{5}\right) \right] \right.  \nonumber \\
&&\times \left\lfloor \left[ \sum_{g,h;e,f}\left( U_{h\delta }^{\dagger
}U_{\gamma g}U_{\beta h}U_{g\alpha }^{\dagger }U_{\delta f}U_{e\gamma
}^{\dagger }\delta _{ef}\right) \right. \right.  \TCItag{27} \\
&&\left. \times Tr\left[ q_{\left( e,\gamma \right) },q_{\left( g,\gamma
\right) };+,c\right] \gamma ^{\nu }\left( 1-\gamma ^{5}\right) \left[
q_{\left( h,\delta \right) },q_{\left( f,\delta \right) };+,c\right] \gamma
^{\mu }\left( 1-\gamma ^{5}\right) \right]  \nonumber \\
&&+\left[ \sum_{k,l;i,j}\left( U_{l\delta }^{\dagger }U_{\gamma k}U_{\delta
j}U_{i\gamma }^{\dagger }U_{\alpha i}U_{j\beta }^{\dagger }\delta
_{kl}\right) \right.  \nonumber \\
&&\left. \left. \left. \times Tr\left[ q_{\left( i,\gamma \right)
},q_{\left( k,\gamma \right) };+,c\right] \gamma ^{\nu }\left( 1-\gamma
^{5}\right) \left[ q_{\left( l,\delta \right) },q_{\left( j,\delta \right)
};+,c\right] \gamma ^{\mu }\left( 1-\gamma ^{5}\right) \right] \right\rfloor
\right\}  \nonumber
\end{eqnarray}%
where one took into account the identity:%
\begin{eqnarray*}
&&Tr\left( M_{1}-\underline{P}_{\left( 1\right) }\right) \gamma _{\nu
}\left( 4z_{3}+\gamma ^{5}\right) \left( M_{2}-\underline{P}_{\left(
2\right) }\right) \gamma _{\mu }\left( 1-\gamma ^{5}\right) \\
&=&Tr\left( M_{1}-\underline{P}_{\left( 1\right) }\right) \gamma _{\nu
}\left( 1-\gamma ^{5}\right) \left( M_{2}-\underline{P}_{\left( 2\right)
}\right) \gamma _{\mu }\left( 4z_{3}+\gamma ^{5}\right)
\end{eqnarray*}%
Of course, one cannot avoid the coherent energy operator expansion according
to (14), and evaluating the traces of gamma matrices one obtains

\begin{eqnarray}
\frac{d\sigma _{\left\{ W,Z\right\} }\left( m\right) }{dy} &=&\frac{d\sigma
_{\left\{ W,Z\right\} }\left( SM\right) }{dy}\left[ 1+\frac{m_{\gamma \gamma
}^{2}}{4}\left( \frac{1}{q_{\left( \gamma \right) }^{02}}+\frac{1}{q_{\left(
\delta \right) }^{02}}\right) \right] -\frac{G^{2}m_{\gamma \delta
}m_{\delta \gamma }\delta _{\alpha \beta }}{2\pi \left\vert \left( P_{\left(
1\right) \cdot }q_{\left( \gamma \right) }\right) \right\vert }  \nonumber \\
&&\times \left\{ \left[ M_{1}M_{2}w_{0}\left( q_{\left( \gamma \right)
}\cdot q_{\left( \delta \right) }\right) +w_{1}\left( P_{\left( 1\right)
\cdot }q_{\left( \gamma \right) }\right) \left( P_{\left( 2\right) \cdot
}q_{\left( \delta \right) }\right) \right] \right.  \nonumber \\
&&\times \left( \frac{\delta _{\alpha \gamma }}{q_{\left( \gamma \right)
}^{02}}+\frac{\delta _{\beta \delta }}{q_{\left( \delta \right) }^{02}}%
\right) +\frac{2\delta _{\alpha \gamma }}{q_{\left( \delta \right) }^{0}}%
\left[ M_{1}M_{2}w_{0}q_{\left( \gamma \right) }^{0}+w_{1}\left( P_{\left(
1\right) \cdot }q_{\left( \gamma \right) }\right) P_{\left( 2\right) }^{0}%
\right]  \TCItag{28} \\
&&\left. +\frac{2\delta _{\beta \delta }}{q_{\left( \gamma \right) }^{0}}%
\left[ M_{1}M_{2}w_{0}q_{\left( \delta \right) }^{0}+w_{1}\left( P_{\left(
2\right) \cdot }q_{\left( \delta \right) }\right) P_{\left( 1\right) }^{0}%
\right] \right\} +O\left( m^{4}\right) ,  \nonumber \\
\frac{d\sigma _{\left\{ W,Z\right\} }\left( SM\right) }{dy} &=&\frac{%
2G^{2}\delta _{\alpha \beta }\delta _{\alpha \gamma }\delta _{\gamma \delta }%
}{\pi \left\vert \left( P_{\left( 1\right) \cdot }q_{\left( \gamma \right)
}\right) \right\vert }\left[ M_{1}M_{2}w_{0}\left( q_{\left( \gamma \right)
}\cdot q_{\left( \delta \right) }\right) +w_{1}\left( P_{\left( 1\right)
\cdot }q_{\left( \gamma \right) }\right) \left( P_{\left( 2\right) \cdot
}q_{\left( \delta \right) }\right) \right] \text{\ \ \ \ \ \ \ \ }  \nonumber
\end{eqnarray}%
Again,the negligible $LIV$ is associated with the neutrino mass while the $%
LI $ Standard Model result is identified with formally zero neutrino mass
limits :

\begin{equation}
\frac{d\sigma _{\left\{ W,Z\right\} }\left( m\right) }{dy}=\frac{d\sigma
_{\left\{ W,Z\right\} }\left( SM\right) }{dy}+O\left( m^{2};LIV\right) 
\tag{29}
\end{equation}%
The overlapping $W-$and \ $Z-$ exchanges cross-section terms of $O\left(
m=0\right) $ are $LI$ \ and flavor conserving, while the negligible terms of 
$O(m^{2})$ carry the $LIV$ terms with, both, the conserved and violate
flavor.

\bigskip

\textbf{\ Discussion---} One thing that one notices right a way is the fact
\ that while the $LIV$ is very real, because it is associated with the $%
O(m^{2})$ terms, it is negligible at least in the scattering-like
experiments. Therefore, the "mass-less" SM is consistent with massive
neutrinos whose masses are $\ \leq 1eV.$ Because they are proportional to $%
O(m^{2})$, the neutrino oscillation scattering cross-sections derived here
are not observable. \ However, the interesting problem to deal with would be
as to how to generalize the negligible neutrino oscillation scattering
cross-sections in the laboratory into the practical long baseline
oscillations probabilities.

\bigskip

\textbf{Acknowledgements---} I wish to thank Dr. Howard E. Brandt for
friendly discussions and help with the computer manipulations. To my wife
Patricia Marie Stone Soln, I am deeply grateful for her expert librarian
help in locating literature over the internet.

\bigskip

\textbf{\ References---}

[1]\qquad Super-KamiokandeCollaboration, Y. Ashie et al., Phys. Rev. Lett. 
\textbf{93}, 101801 (2004); M. Shiozawa, Prog. Part. Nucl. Phys. \textbf{57}%
, 79 (2006).

[2]\qquad SNO Collaboration, Phys. Rev. Lett. \textbf{81,} 071301 (2001); 
\textbf{89}, 011301 (2002); \textbf{89,} 011302 (2002); Phys. Rev. C\textbf{%
72}, 055502 (2005).

[3]\qquad KAMLAND Collaboration, T. Araki et al., Phys. Rev. Lett.\textbf{94}%
, 081801 (2005).

[4]\qquad Homestake Collaboration, T. Leveland et al. Astrophys. J. \textbf{%
496,} 505 (1998); GNO Collaboration, M. Altman et al. Phys. Lett. B\textbf{%
616}, 174 (2005); SAGE Collaboration, J. N. Abdurashitov et al., \ \ \ \ \ \
\ \ \ \ \ Nucl. Phys. Proc. Suppl. \textbf{110}, 315 (2002);
Super-KamiokandeCollaboration, J. Hosaka et al., Phys. Rev. D\textbf{73},
112001 (2006).

[5] S. M. Bilenky, C. Giunti and W. Grimus, Progress In Particle and Nuclear
Physics, \textbf{43}, 1 (1999).

[6] C. Giunti and M. Laveder, \textquotedblleft Neutrino
Mixing\textquotedblright ; hep-ph/0310238v2.

[7] B. Kayser, \textquotedblleft Neutrino Oscillation
Phenomenology\textquotedblright ; Proc. of the 61 st Scottish Universities
Summer School in Physics, Eds. C. Frogatt and P. Soler (to appear); arXiv:
0804.1121v3 [hep-ph].

[ [8] M. Fukugita and T. Yanagida, \textquotedblleft Physics of Neutrinos
and Applications to Astrophysics\textquotedblright\ (Springer Verlag Berlin
Heidelberg 2003).

[9] C. Giunti and C. W. Kim, \textquotedblleft Fundamentals of Neutrino
Physics and Astrophysics\textquotedblright\ (Oxford University Press, Oxford
2007).

[10] C. Giunti, \textquotedblleft Neutrino Flavor States and the Quantum
Theory of Neutrino Oscillations\textquotedblright\ (XI Mexican Workshop on
Particles and Fields, 7-12 November 2007)), arXiv: 0801. 0653 v1 [hep-ph].

11] C. Giunti, \textquotedblleft Fock States of Flavor Neutrinos are
Unphysical\textquotedblright ; Eur. Phys. J. C\textbf{39,} 377-382 (2005);
hep- ph/0312256v2.

[12] S. M. Bilenky, F. von Feilitzsch and W. Potzel, J. Phys. G; Nucl. Part.
Phys. \textbf{34}, 987 (2007), hep- ph/0611285v2.

.

\bigskip

\bigskip

\ \ \ \ \ \ \ \ \ \ \ \ \ \ \ \ \ \ \ \ \ \ \ \ \ \ \ \ \ \ \ \ \ \ \ \ \ \
\ \ \ \ \ \ \ \ \ \ \ \ \ \ \ \ \ \ \ \ \ \ \ \ \ \ \ \ \ \ \ \ \ \ \ \ \ \
\ \ \ \ \ \ \ \ \ \ \ \ \ \ \ \ \ \ \ \ \ \ \ \ \ \ \ \ \ \ \ \ \ \ \ \ \ \
\ \ \ \ \ \ \ \ \ \ \ \ \ \ \ \ \ \ \ \ \ \ \ \ \ \ \ \ \ \ \ \ \ \ \ \ \ \
\ \ \ \ \ \ \ \ \ \ \ \ \ \ \ \ \ \ \ \ \ \ \ \ \ \ \ \ \ \ \ \ \ \ \ \ \ \
\ \ \ \ \ \ \ \ \ \ \ \ \ \ \ \ \ \ \ \ \ \ \ \ \ \ \ \ \ \ \ \ \ \ \ \ \ \
\ \ \ \ \ \ \ \ \ \ \ \ \ \ \ \ \ \ \ \ \ \ \ \ \ \ \ \ \ \ \ \ \ \ \ \ \ \
\ \ \ \ \ \ \ \ \ \ \ \ \ \ \ \ \ \ \ \ \ \ \ \ \ \ \ \ \ \ \ \ \ \ \ \ \ \
\ \ \ \ \ \ \ \ \ \ \ \ \ \ \ \ \ \ \ \ \ \ \ \ \ \ \ \ \ \ \ \ \ \ \ \ \ \
\ \ \ \ \ \ \ \ \ \ \ \ \ \ \ \ \ \ \ \ \ \ \ \ \ \ \ \ \ \ \ \ \ \ \ \ \ \
\ \ \ \ \ \ \ \ \ \ \ \ \ \ \ \ \ \ \ \ \ \ \ \ \ \ \ \ \ \ \ \ \ \ \ \ \ \
\ \ \ \ \ \ \ \ \ \ \ \ \ \ \ \ \ \ \ \ \ \ \ \ \ \ \ \ \ \ \ \ \ \ \ \ \ \
\ \ \ \ \ \ \ \ \ \ \ \ \ \ \ \ \ \ \ \ \ \ \ \ \ \ \ \ \ \ \ \ \ \ \ \ \ \
\ \ \ \ \ \ \ \ \ \ \ \ \ \ \ \ \ \ \ \ \ \ \ \ \ \ \ \ \ \ \ \ \ \ \ \ \ \
\ \ \ \ \ \ \ \ \ \ \ \ \ \ \ \ \ \ \ \ \ \ \ \ \ \ \ \ \ \ \ \ \ \ \ \ \ \
\ \ \ \ \ \ \ \ \ \ \ \ \ \ \ \ \ \ \ \ \ \ \ \ \ \ \ \ \ \ \ \ \ \ \ \ \ \
\ \ \ \ \ \ \ \ \ \ \ \ \ \ \ \ \ \ \ \ \ \ \ \ \ \ \ \ \ \ \ \ \ \ \ \ \ \
\ \ \ \ \ \ \ \ \ \ \ \ \ \ \ \ \ \ \ \ \ \ \ \ \ \ \ \ \ \ \ \ \ \ \ \ \ \
\ \ \ \ \ \ \ \ \ \ \ \ \ \ \ \ \ \ \ \ \ \ \ \ \ \ \ \ \ \ \ \ \ \ \ \ \ \
\ \ \qquad

\end{document}